\documentclass[lettersize,journal,onecolumn]{IEEEtran} %1
\usepackage{amsmath,amsfonts}
\usepackage{algorithmic}
\usepackage{array}
\usepackage[caption=false,font=normalsize,labelfont=sf,textfont=sf]{subfig}
\usepackage{textcomp}
\usepackage{stfloats}
\usepackage{url}
\usepackage{verbatim}
\usepackage{multirow}
\usepackage{ulem}
\usepackage{booktabs}
\usepackage{graphicx,subfig}
\usepackage{mathptmx}
\def\BibTeX{{\rm B\kern-.05em{\sc i\kern-.025em b}\kern-.08em
		T\kern-.1667em\lower.7ex\hbox{E}\kern-.125emX}}
\usepackage{balance}
\begin{document}
	\title{Multimodal Graph Neural Network for Recommendation with Dynamic De-redundancy and Modality-Guided Feature De-noisy}
	\author{Feng Mo, Lin Xiao, Qiya Song, Xieping Gao, and Eryao Liang
		\thanks{F. Mo, L. Xiao, and Q. Song are with Hunan Provincial Key Laboratory of Intelligent Computing and Language Information Processing, Hunan Normal University}
	}
	
	\markboth{Journal of \LaTeX\ Class Files,~Vol.~18, No.~9, September~2020}%
	{How to Use the IEEEtran \LaTeX \ Templates}
	
	\maketitle
	
	\begin{abstract}
		Graph neural networks (GNNs) have become crucial in multimodal recommendation tasks because of their powerful ability to capture complex relationships between neighboring nodes. However, increasing the number of propagation layers in GNNs can lead to feature redundancy, which may negatively impact the overall recommendation performance. In addition, the existing recommendation task method directly maps the preprocessed multimodal features to the low-dimensional space, which will bring the noise unrelated to user preference, thus affecting the representation ability of the model. To tackle the aforementioned challenges, we propose Multimodal Graph Neural Network for Recommendation (MGNM) with Dynamic De-redundancy and Modality-Guided Feature De-noisy, which is divided into local and global interaction. Initially, in the local interaction process,we integrate a dynamic de-redundancy (DDR) loss function which is achieved by utilizing the product of the feature coefficient matrix and the feature matrix as a penalization factor. It reduces the feature redundancy effects of multimodal and behavioral features caused by the stacking of multiple GNN layers. Subsequently, in the global interaction process, we developed modality-guided global feature purifiers for each modality to alleviate the impact of modality noise. It is a two-fold guiding mechanism eliminating modality features that are irrelevant to user preferences and captures complex relationships within the modality. Experimental results demonstrate that MGNM achieves superior performance on multimodal information denoising and removal of redundant information compared to the state-of-the-art methods.
	\end{abstract}
	
	
	\section{Introduction}
	\IEEEPARstart{M}{ultimodal} recommendation methods (MRs) are essential in the field of Artificial Intelligence (AI) and are widely applied in scenarios, i.e., healthcare \cite{drug1,drug2}, online platform \cite{os1,os2,os3}, news recommendation \cite{news1,news2}, and social media \cite{social1,social2}. It uses multimodal signals to bootstrap collaborative signals to better capture the user's interests and provide recommendations that are better suited to the user's needs. Current MRs typically consist of three main steps: multimodal information fusion, collaborative filtering, and prediction.
	\begin{figure}[h]
		\centering
		\includegraphics[width=0.5\textwidth]{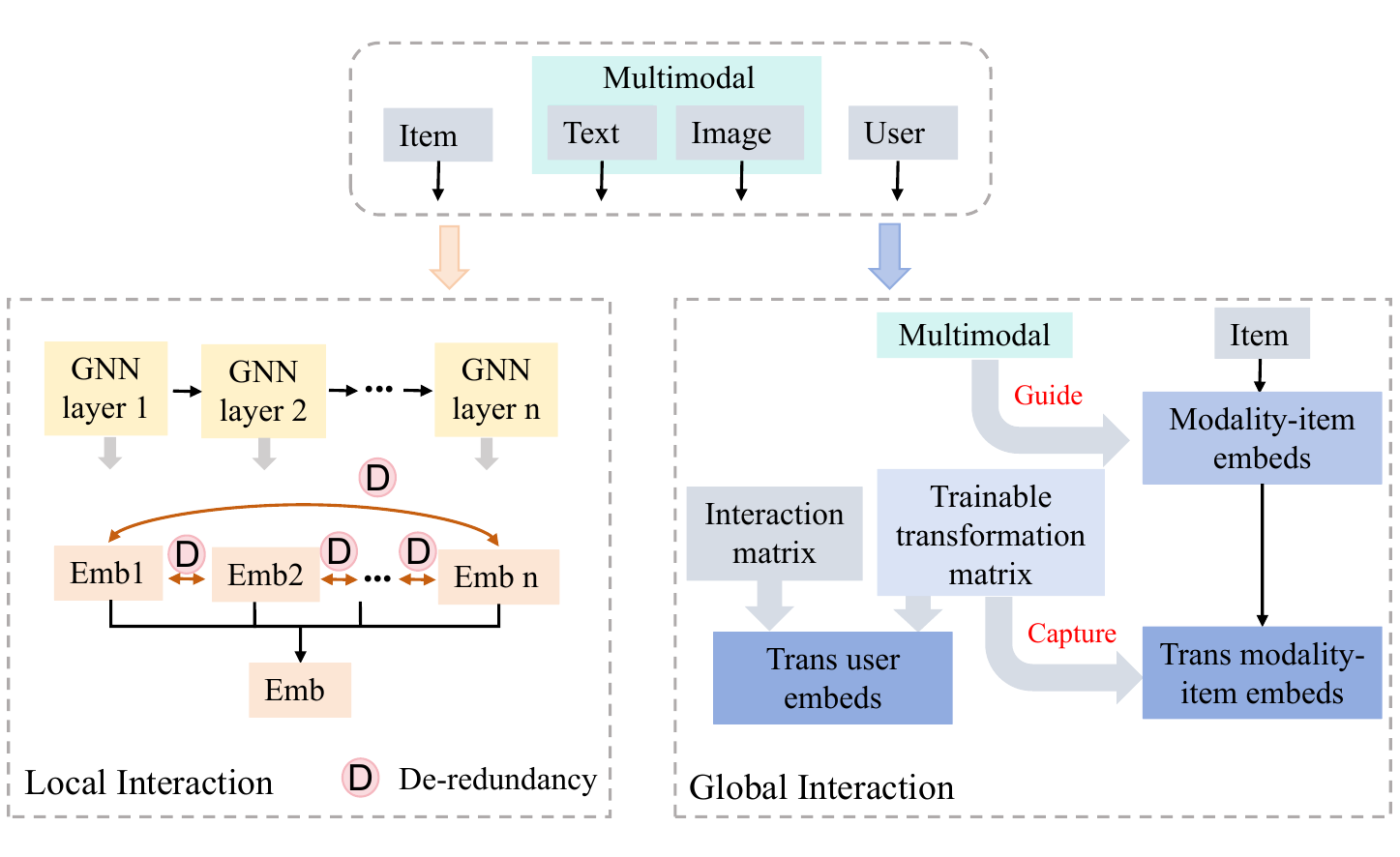}
		\caption{Illustration of the proposed model. It is structured into global and local interaction. Firstly, global interaction leverages modality features to guide the elimination of redundant information irrelevant to user preferences. Secondly, a trainable transformation matrix captures complex relationships within the modality, providing global preference information. For local interaction, the model addresses feature redundancy from GNNs by penalizing over correlated features. Predictions are made by calculating the weighted sum of global and local interaction information. This approach captures both global and local user interests, offering a comprehensive characterization of user preferences.}
		\label{fuzhu}
	\end{figure}
	
	Early MRs often linearly integrate individual modality signals into collaborative signals to establish user and item representation models. These methods primarily address low-order user-item interactions, neglecting higher-order relationship modeling. Additionally, different users have different preferences for various modalities. Given the inherent advantage of graph neural networks (GNNs) in capturing neighbor signals, applying GNNs to MRs is a natural choice. Specifically, GNNs aggregate neighbor information to capture more interaction details. 
	Despite the commendable performance achieved by existing models, they still struggle with the following challenges: 
	Many existing GNNs-based recommendation models contain only two layers of GNNs, this is because the effect of extracting interactive behavioral features or multimodal features does not keep going up as the number of GNNs layers continue to be stacked. Some researchers ascribe this phenomenon to the over-smoothing problem \cite{guopinghua1, guopinghua2, guopinghua3,guopinghua4}, where node representations gradually become similar with the increase in layers. But Wu et al. \cite{AFDGCF} argue that as the number of GNNs layers increases, excessive feature correlations emerge, leading to performance degradation. This is manifested in the increasing redundancy among node representation features, which adversely affects the quality of learned embeddings. 
	Existing models typically preprocess multimodal information through pretraining and then directly map embeddings to a smaller dimensionality \cite{mgcn, bm3, mmgcn}. However, directly mapping modality information from a high dimensional space to a low dimensional space will retain unnecessary noise information with greater probability. The introduction of modality noise information can directly affect the quality of the embedding and ultimately the expression of user preferences.

	\section{Related Work} 
	\label{related work}
	In this section, we introduce related work from three perspectives: traditional recommendation, single-modal GNNs for recommendation and multimodal GNNs for recommendation.
	
	\subsection{Traditional GNNs Recommendation}
	In the field of MRs, Collaborative Filtering (CF) based on Graph Neural Networks (GNNs) has become increasingly popular. Unlike traditional matrix factorization (MF) \cite{mf1, mf2} methods, GNNs-based CF leverages the inherent higher-order connections in user historical interaction data to enhance representation performance. Initially, some studies directly adopted the information propagation mechanism of ordinary GNNs, aggregating higher-order neighbor information to represent users and items \cite{REGCMC, GraphCM, GNN1, mmgcn,graph1,graph2}. With deeper research into GNNs, many researchers have further improved recommendation performance by simplifying the message propagation process of GNNs \cite{UltraGCN, Revisiting, LightGCN}. 
	According to CAGCN \cite{CAGCN}, the message passing mechanism might not be sufficient in utilizing collaboration signals to accurately predict user preferences, leading to the incorporation of a selective aggregation mechanism.
	With the rapid rise of attention mechanisms, an increasing number of studies are inclined to incorporate attention mechanisms into GNNs to enhance user-item representations \cite{Graphatt1, Graphatt2}.
	
	\subsection{Single-modal GNNs for Recommendation}
	With the development of multimodal learning, more and more researchers commit to integrate modality signals with collaborative signals for recommendation. Based on the type of modalities used, it can be categorized into the following two types, i.e., visual modeling, textual modeling.
	
	\subsubsection{Visual Modeling}
	MRs based on visual modality are typically suitable for scenarios highly dependent on visual signals. A classic example is VBPR \cite{VBPR}, which enriches item representations by simply concatenating pre-trained visual features. With the widespread use of Convolutional Neural Network (CNN), subsequent studies \cite{CNN1, CNN2} have attempted to integrate CNN into recommendation frameworks, effectively improving model performance. However, many studies indicate that visual information has less impact on recommendation effectiveness compared to textual information%\cite{}
	. This is due to semantic differences between images and text, making it challenging for the model to fully perceive visual features during representation. Therefore, solely relying on visual information may not construct effective user preferences.
	
	\subsubsection{Textual Modeling}
	In the realm of Recommender Systems (RSs), common textual modalities include item descriptive titles and user comments. 
	A classic method is DeepCoNN \cite{DeepCoNN}, where one network focuses on learning user behavior from user comments, while another network learns item attributes from comments written for that item. 
	By coupling the last layer of two parallel neural networks, DeepCoNN jointly learns item attributes and user behavior from textual comments. With the popularity of attention mechanisms, some scholars utilize attention mechanisms to calculate combined interaction information \cite{attention1} and the importance of each word in a comment \cite{attention2}. However, with the evolution of MRs, single modality recommendation tends to perform less effectively compared to multimodal recommendation. This is because multimodal recommendation integrates more auxiliary information from different modalities, enhancing the robustness of the model.
	
	\section{Problem Description}
	\label{Problem Description}
	We denote the user set as $U=\{u_{1}, u_{2},...,u_{p}\}$ and the item set as $I=\{i_{1}, i_{2},...,i_{q}\}$. For each \( u \in U \) and \( i \in I \), their embeddings are represented as $e_{u} \in \mathbb{R} ^{d}$  and  $e_{i} \in \mathbb{R} ^{d}$, respectively, where $d$ represent the dimensions of user and item embeddings, respectively. The interaction between users and items is represented as a matrix $ R \in \left \{ 0,1  \right \} ^{p\times q}$, where each element $ r_{u,i}\in  \mathbb{R} $ indicates the interaction behavior between a user $ u $ and an item $ i $, i.e., clicks or purchases. A value of 1 indicates that the interaction has occurred, while 0 indicates no interaction. In addition to user-item interactions, each item is also associated with multimodal content information $ M\in \left \{ v,t \right \} $, where $ v $ and $ t $ represent visual and textual features, respectively. The multimodal information of each item $ i $ is represented as $ e_{i}^{m}\in  \mathbb{R} ^{d_{m} } $, where \( d_m \) represents the feature dimension of the modality $ m $. In this article, we only consider two mainstream modalities: text $ t $ and vision $ v $.
	
	Based on the above settings, multimodal recommendation aims to learn a prediction function that predicts the final score $ $ between user $ u $ and item $ i $ by jointly modeling user behavior and multimodal content. This can be represented as:
	\begin{equation}
		\hat{r} _{u,i}=Pred(R,E^{id},\left \{ E_{i}^{m}  \right \}_{m\in M} ) ,
	\end{equation}
	where $ Pred\left (  \cdot  \right )  $ denotes the prediction function, $ E^{id}=\left [ e_{u1},...,e_{up},e_{i1},...,e_{iq} \right ] \in \mathbb{R} ^{\left ( p+q \right )\times d } $ denotes the embedding matrix stacked with all user ID embeddings and item ID embeddings, and $ E_{i}^{m}=\left [ e_{i1}^{m},..., e_{iq}^{m}\right ] \in \mathbb{R} ^{q \times d } $ denote the matrix of item multimodal features under modality $ m $. 
	
	\section{Proposed Method}
	\label{Proposed Method}
	MGNM primarily consists of three parts: Local Interaction, Global Interaction, and Prediction. Firstly, the Local Interaction part is divided into User-Item-based GNNs Collaborative Filtering and Modality-based GNNs Collaborative Filtering. By inputting the feature after embedding into GNNs for encoding and then weighting the results, local representations of users and items with modality features are obtained. Secondly, the original embedding information is transformed into a low-dimensional feature through the global feature filter based on modality features. Through Global Interaction, modality information is integrated into global information to obtain global representations containing modality representations. Finally, the Local Interaction and Global Interaction are combined and predicted by weighting and fusing the final user and item outputs.
	
	\subsection{Local Interaction}
	Local Interaction mainly learns collaborative-related and modality-related user-item representations.

	We first capture the high-order connections of users and items through the message propagation mechanism on the graph. Specifically, the collaborative graph propagation function $ Prop\left ( \cdot  \right )  $ at layer $ l+1 $ can be represented as:
	\begin{equation}
		E^{l+1}=Prop\left ( E^{l}  \right )  =\left ( D^{-\frac{1}{2} }A D^{-\frac{1}{2} } \right ) E^{l} ,
	\end{equation}
	where $ Prop\left ( \cdot  \right ) $ represents the lightweight simplified graph convolutional neural network \cite{LightGCN}. $ A\in \mathbb{R} ^{\left ( p + q \right )\times \left ( p + q \right )} $ represents the adjacency matrix constructed from the interaction matrix $ R $, and $ D $ represents the degree matrix of $ A $, where each diagonal element $  $ indicates the number of non-zero elements in the corresponding row $ t $ of the matrix $ A $. The initial embedding matrix is set as $ E^{0}=E^{id} $, and then the results of all hidden layers are combined as follows:
	\begin{equation}
		E_{loc}^{id}=Comb\left ( E^{0} ,...,E^{l}\right )   ,
	\end{equation}
	where $ E_{loc}^{id}\in \mathbb{R}^{\left ( p+ q \right )\times d }  $ represents the collaborative-related embeddings of local information of users and items, and $ Comb\left ( \cdot  \right )  $ denotes the combination between layers.
	
	Due to semantic differences between different modalities, the impact of different modalities on user-item representations varies. We further infer modality-related embeddings of users and items in the interaction graph with modality features. The original modality features of items are usually generated by pre-trained models, i.e., ResNet50 \cite{resnet50}, BERT \cite{bert}, and ViT \cite{vit}, which have different dimensions in different feature spaces. In this article, we utilize the processing methods provided by MMRec \footnote[1]{https://github.com/enoche/MMRec} for extracting modality features.
	
	We map the high-dimensional modality features $ e_{i}^{m} $ of each item to a unified embedding space $  \mathbb{R} ^{d} $, as shown below:
	\begin{equation}
		\hat{e} _{i}^{m}=Trans\left ( e_{i}^{m} \right ) = e_{i}^{m}\cdot W_{m}    ,
	\end{equation}
	we transform the item $ i $ modality embeddings into modality features through a transformation function $  Trans\left ( \cdot  \right ) $, where $  Trans\left ( \cdot  \right ) $ is a mapping function defined by transformation matrix $ W_{m}\in  \mathbb{R}^{d_{m} \times d}  $. This transformation allows mapping high-dimensional modality features to a lower-dimensional space. Then, we initialize the user modality features by aggregating the modality features of items. Specifically, we perform matrix multiplication between the adjacency matrix $ A $ and the modality features of items. For the obtained features, we apply a penalty factor, where the weight of features decreases as the number of interacting items increases, as shown below:
	\begin{equation}
		\hat{e} _{u}^{m} = \frac{1}{\left | N_{u}  \right | } {\textstyle \sum_{i\in N_{u}}} \hat{e} _{i}^{m}    ,
	\end{equation}
	where $ N_{u} $ represents the set of neighbors of user $ u\in U $ in the interaction graph $ G $. Finally, the obtained modality-related user and item representations are combined together:
	\begin{equation}
		\hat{E}^{m}=\left [ \hat{e}_{u_{1} }^{m},...,\hat{e}_{u_{p} }^{m}, \hat{e}_{i_{1} }^{m},...,\hat{e}_{i_{p} }^{m} \right ] \in R^{\left ( p+q \right )\times d} ,
	\end{equation}
	the representation obtained is used as the initial input for modality graph propagation:
	\begin{equation}
		\hat{E}^{m,k+1}=Prop_{mm}\left ( \hat{E}^{m,k} \right ) =\left ( D^{-\frac{1}{2} }A D^{-\frac{1}{2} } \right )\hat{E}^{m,k}.
	\end{equation}
	
	We select the high-order modality embedding, specifically the $ k $ layer, as the modality-related embedding, denoted as $ E_{loc}^{m} =\hat{E}^{m,k} $. Finally, we normalize the collaborative-related user and item representations and modality-related user and item representations, and then compute the weighted sum to obtain the local user and item representations $ E_{loc}$:
	\begin{equation}
		E_{loc}=E_{loc}^{id}+Form\left ( \sum_{m\in M} E_{loc}^{m}  \right ) ,
	\end{equation}
	where $ Form\left ( \cdot   \right )  $ represents the normalization function.
	
	\subsection{Global Interaction}
	Items are enriched with meaningful content through modality information, but it may also introduce modality noise.
	We design a global feature filter based on modality features to filter out modality noise guided by item information. Specifically, we first map the original modality features to a moderate dimension, then transform these moderate-dimensional features into high-order features:
	\begin{equation}
		\dot{E}_{i}^{m}=W_{1}E_{i}^{m}+b_{1},
	\end{equation}
	where $ W_{1}\in R^{4d\times d_{m} } $ and $ b_{1} \in R^{d}$ represent trainable transformation matrices and bias vectors, respectively. Then, a gating mechanism is applied to map them to a low dimensional space, thus filtering out some noise:
	\begin{equation}
		\ddot{E}_{i}^{m}=gate\left ( E^{id} \dot{E}_{i}^{m}\right )=E^{id}\odot \sigma \left ( W_{2}\dot{E}_{i}^{m}+b_{2}   \right ),
	\end{equation}
	in this equation, $ W_{2}\in R^{d\times 4d}  $ and $ b_{2} \in R^{d}$ are both trainable parameters. $ \odot$ represents element-wise multiplication, and $ \sigma$ is the sigmoid function.
	
	Then, we define a learnable transformation vector $ \left \{ t_{a}^{m}  \right \} _{a=1}^{B} \left ( t_{a}^{m} \in R^{^{d_{m} } }  \right ) $, through which the modelity can adaptively learn the implicit dependencies between modality information and users/items. Here, $ B$ represents the dimension of the transformation vector, which is configured to 4 in our experiments. By integrating the transformation vector with the filtered modality features, we obtain the project representation information containing modality features $ H_{i}^{m} $:
	\begin{equation}
		H_{i}^{m}=\ddot{E}_{i}^{m}  \cdot T^{m^{T} } .
	\end{equation}
	
	We perform feature-level fusion of $ H_{i}^{m}$ with the adjacency matrix $A$ to obtain the user representation information containing modality features $ H_{u}^{m}$:
	\begin{equation}
		H_{u}^{m}= A  \cdot T^{m^{T} } ,
	\end{equation}
	where $ T^{m} =\left [ t_{1}^{m},...,t_{B}^{m}  \right ] \in R^{B\times d_{m} } $ represents the matrix composed of transformation vectors. 
	
	By computing the propagation of $ \tilde{H} _{i}^{m} $ and $ \tilde{H} _{u}^{m} $ on the transformation matrix, we obtain the final global user and item representations $ E_{glo}$:
	\begin{equation}
		E_{i}^{m,f+1}=Drop\left ( \tilde{H}_{i}^{m}\cdot  \tilde{H}_{i}^{m^{T} }  \right )  \cdot E_{i}^{m,f},
	\end{equation}
	\begin{equation}
		E_{u}^{m,f+1}=Drop\left ( \tilde{H}_{u}^{m}\cdot  \tilde{H}_{i}^{m^{T} }  \right )  \cdot E_{i}^{m,f},
	\end{equation}
	\begin{equation}
		E_{glo}=\sum_{u\in U} \left [ E_{u}^{m,H},E_{i}^{m,H} \right ].
	\end{equation}
	Similar to LGMRec \cite{lgmrec}, to further achieve robust fusion of global embeddings across different modalities, we utilize cross-modal contrastive learning to extract self-supervised signals for achieving global interest consistency. The cross-modal contrastive loss on the user side is defined as follows:
	\begin{equation}
		L_{u}^{HCL}=\sum_{u\in U}-log\frac{exp\left ( s\left ( E_{u}^{v,H}, E_{u}^{t,H} \right )/\tau   \right ) }{ {\textstyle \sum_{u^{'}\in U }} exp\left ( s\left ( E_{u}^{v,H}, E_{u}^{t,H} \right )/\tau   \right )} ,
	\end{equation}
	where $s\left ( \cdot  \right ) $ is the cosine similarity function and $\tau$ is the temperature vector, set to 0.2. Similarly, we compute the cross-modal contrastive loss $L_{i}^{HCL}$ on the item side. 
	
	Finally, obtain the final representation $ E^{*} $ by fusing the local embeddings of users and items $E_{loc}$ and the global embeddings $E_{glo}$:
	\begin{equation}
		E^{*} =E_{loc}+\sum_{m\in M}\alpha \cdot Norm\left ( E_{glo} \right )  ,
	\end{equation}
	where $ \alpha$ is a tunable parameter used to control the global embeddings.
	
	\subsection{Prediction}
	To train our model, we employ the Bayesian Personalized Ranking (BPR) loss \cite{bpr} to evaluate users' preference scores for items. BPR is a pairwise loss function that promotes higher scores for observed user-item pairs compared to unobserved ones:
	\begin{equation}
		R=\left \{ \left ( u,i,j \right )|\left ( u,i \right ) \in \varepsilon ,\left ( u,j \right ) \notin \varepsilon  \right \}   ,
	\end{equation}
	\begin{equation}
		L_{bpr}=\sum_{\left ( u,i,j \right )\in R }-ln\sigma \left ( r_{ui} - r_{uj}\right )  +\lambda \left \| \Theta  \right \|_{2}^{2},
	\end{equation}
	where $ R$ represents the triplet used for training, $ \sigma \left ( \cdot  \right ) $ is a sigmoid function, and $ \lambda$ and $ \Theta$ are regularization coefficients and model parameters, respectively.
	
	To alleviate excessive correlation in the feature dimensions learned by the model for users and items, we employ a loss \cite{AFDGCF} term to constrain the correlation between features. First, we compute the column-wise correlations of the matrix $ p_{cov}$. Since the correlation of a vector with itself is always 1, we define a matrix of column correlations $ P_{R}$ to represent it:
	\begin{equation}
		p_{cov}=\sqrt{\frac{1}{2} \left \| P_{R}-I_{d} \right \|_{F}^{2}  } =\frac{1}{\sqrt{2} }\left \| P_{R}-I_{d} \right \|,
	\end{equation}
	where $ I_{d}$ is the identity matrix with $d\times d $, and the column correlation matrix $ P_{R}$ can be computed as follows: 
	\begin{equation}
		P_{R_{ij} }=\frac{Cov\left ( R_{ij} \right ) }{\sqrt{Cov\left ( R_{ii} \right )Cov\left ( R_{jj} \right )} },
	\end{equation}
	where $Cov\left ( R \right )$ represents the covariance matrix among the column vectors of the matrix:
	\begin{equation}
		Cov\left ( R \right )  =\left ( E_{id}^{loc}-\tilde{E}_{id}^{loc}   \right ) ^{T}\left ( E_{id}^{loc}-\tilde{E}_{id}^{loc}   \right ),
	\end{equation}
	where $\tilde{E}_{id}^{loc}$ represents the matrix composed of the mean values of each column of $E_{id}^{loc}$. By separately calculating the correlation coefficient matrices for users and items, we obtain the optimization objective $L_{afd}$:
	\begin{equation}
		L_{ddr}=\sum_{l=0}^{L} \left [ \mu _{U}^{l}p_{cov}\left ( E_{loc}^{uid^{l}}  \right ) +\mu _{I}^{l}p_{cov}\left ( E_{loc}^{iid^{l}}  \right )   \right ],
	\end{equation}
	where $\mu _{U}^{l}$ and $\mu _{I}^{l}$ respectively represent the penalty coefficients for the $l$-th layer corresponding to users and items. The specific calculation process is as follows:
	\begin{equation}
		\mu _{U}^{l}=\frac{1/p_{cov}\left ( E_{loc}^{uid^{l}}  \right ) }{ {\textstyle \sum_{i=0}^{L}1/p_{cov}\left ( E_{loc}^{uid^{i}}  \right ) } },
	\end{equation}
	\begin{equation}
		\mu _{I}^{l}=\frac{1/p_{cov}\left ( E_{loc}^{iid^{l}}  \right ) }{ {\textstyle \sum_{i=0}^{L}1/p_{cov}\left ( E_{loc}^{iid^{i}}  \right ) } },
	\end{equation}
	similarly, we compute the loss for modality-related user-item pairs $L_{ddr_{mm}}$. 
	
	Finally, we obtain the ultimate loss function $L$:
	\begin{equation}
		L=L_{bpr}+\omega \cdot \left ( L_{HCL}^{u} +L_{HCL}^{i} \right )  +\beta \cdot L_{ddr}+\delta \cdot L_{ddr_{mm}} ,
	\end{equation}
	Where $\omega $ represents the hyperparameter for weighting the loss terms, $ \beta$  and $ \delta$ are hyperparameters used to adjust the relative magnitude of the feature dynamic de-redundancy loss. Detailed tuning will be provided in Section \ref{hyperparameter}.
	
	\section{Experiments}
	\label{Experiments}
	In this section, we present the experiments performed to evaluate the performance of our proposed approach on sample datasets. 
	We examine the results obtained by varying parameters and techniques within the system and compare these findings with those from other cutting-edge models in the realm of multimodal recommender systems.
	
	% \usepackage{multirow}
	% \usepackage[normalem]{ulem}
	\begin{table*}[t!]
		\caption{Performance comparison of the baselines and our model in terms of Recall and NDCG over the three datasets. We emphasize the leading outcomes across the globe for each dataset and metric by using bold font, with the second-best results being underlined.}
		\centering
		\label{table2}
		\begin{tabular}{c|c|cc|ccccccc|c}
			\hline
			\multirow{2}{*}{Dataset} & \multirow{2}{*}{Metrics} & \multicolumn{2}{c|}{Traditional Methods} & \multicolumn{7}{c|}{Multimodal Methods}                                            & \multirow{2}{*}{Improve} \\ \cline{3-11}
			&                          & BPR             & LightGCN          & VBPR   & MMGCN  & DualGNN & LATTICE      & BM3    & MGCN         & MGNM             &                          \\ \hline
			\multirow{8}{*}{Baby}    & Recall@5                 & 0.0165          & 0.0238            & 0.0239 & 0.0245 & 0.0311  & 0.0336       & 0.0329 & {\uline{0.0393}} & \textbf{0.0412} & 4.83\%                   \\
			& Recall@10                & 0.0268          & 0.0402            & 0.0397 & 0.0397 & 0.0518  & 0.0535       & 0.0539 & {\uline{0.0608}} & \textbf{0.0666} & 9.54\%                   \\
			& Recall@20                & 0.0441          & 0.0644            & 0.0665 & 0.0641 & 0.0820  & 0.0830       & 0.0848 & {\uline{0.0927}} & \textbf{0.1003} & 8.20\%                   \\
			& Recall@50                & 0.0823          & 0.1140            & 0.1210 & 0.1185 & 0.1419  & 0.1445       & 0.1480 & {\uline{0.1554}} & \textbf{0.1669} & 7.40\%                   \\
			& NDCG@5                   & 0.0110          & 0.0158            & 0.0158 & 0.0156 & 0.0204  & 0.0222       & 0.0214 & {\uline{0.0263}} & \textbf{0.0275} & 4.56\%                   \\
			& NDCG@10                  & 0.0144          & 0.0274            & 0.0210 & 0.0206 & 0.0273  & 0.0287       & 0.0283 & {\uline{0.0333}} & \textbf{0.0358} & 7.51\%                   \\
			& NDCG@20                  & 0.0188          & 0.0375            & 0.0279 & 0.0269 & 0.0350  & 0.0363       & 0.0362 & {\uline{0.0415}} & \textbf{0.0445} & 7.23\%                   \\
			& NDCG@50                  & 0.0266          & 0.0053            & 0.0389 & 0.0379 & 0.0471  & 0.0488       & 0.0491 & {\uline{0.0542}} & \textbf{0.0579} & 6.83\%                   \\ \hline
			\multirow{8}{*}{Video}   & Recall@5                 & 0.0453          & 0.0543            & 0.0745 & 0.0511 & 0.0780  & 0.0775       & 0.0751 & {\uline{0.0849}} & \textbf{0.0915} & 7.77\%                   \\
			& Recall@10                & 0.0722          & 0.0873            & 0.1198 & 0.0843 & 0.1200  & 0.1234       & 0.1166 & {\uline{0.1345}} & \textbf{0.1408} & 4.68\%                   \\
			& Recall@20                & 0.1106          & 0.1351            & 0.1796 & 0.1323 & 0.1807  & 0.1855       & 0.1772 & {\uline{0.1997}} & \textbf{0.2056} & 2.95\%                   \\
			& Recall@50                & 0.1885          & 0.2258            & 0.2847 & 0.2277 & 0.2848  & 0.2937       & 0.2855 & {\uline{0.3132}} & \textbf{0.3163} & 0.99\%                   \\
			& NDCG@5                   & 0.0298          & 0.0366            & 0.0498 & 0.0330 & 0.0518  & 0.0524       & 0.0499 & {\uline{0.0577}} & \textbf{0.0625} & 8.32\%                   \\
			& NDCG@10                  & 0.0386          & 0.0475            & 0.0647 & 0.0440 & 0.0656  & 0.0675       & 0.0636 & {\uline{0.0740}} & \textbf{0.0788} & 6.49\%                   \\
			& NDCG@20                  & 0.0486          & 0.0599            & 0.0802 & 0.0565 & 0.0814  & 0.0837       & 0.0793 & {\uline{0.0910}} & \textbf{0.0957} & 5.16\%                   \\
			& NDCG@50                  & 0.0645          & 0.0785            & 0.1019 & 0.0761 & 0.1028  & 0.1060       & 0.1015 & {\uline{0.1144}} & \textbf{0.1185} & 3.58\%                   \\ \hline
			\multirow{8}{*}{Beauty}  & Recall@5                 & 0.0308          & 0.0428            & 0.0487 & 0.0332 & 0.0599  & {\uline{0.0636}} & 0.0543 & 0.0572       & \textbf{0.0698} & 9.75\%                   \\
			& Recall@10                & 0.0512          & 0.0675            & 0.0763 & 0.0546 & 0.0897  & {\uline{0.0962}} & 0.0842 & 0.0862       & \textbf{0.1051} & 9.25\%                   \\
			& Recall@20                & 0.0735          & 0.0985            & 0.1108 & 0.0871 & 0.1308  & {\uline{0.1393}} & 0.1209 & 0.1241       & \textbf{0.1505} & 8.04\%                   \\
			& Recall@50                & 0.1150          & 0.1540            & 0.1705 & 0.1439 & 0.1988  & {\uline{0.2092}} & 0.1875 & 0.1850       & \textbf{0.2236} & 6.88\%                   \\
			& NDCG@5                   & 0.0202          & 0.0284            & 0.0325 & 0.0218 & 0.0407  & {\uline{0.0428}} & 0.0355 & 0.0392       & \textbf{0.0469} & 9.58\%                   \\
			& NDCG@10                  & 0.0271          & 0.0366            & 0.0417 & 0.0290 & 0.0506  & {\uline{0.0536}} & 0.0455 & 0.0488       & \textbf{0.0587} & 9.51\%                   \\
			& NDCG@20                  & 0.0331          & 0.0448            & 0.0508 & 0.0375 & 0.0614  & {\uline{0.0649}} & 0.0551 & 0.0589       & \textbf{0.0706} & 8.78\%                   \\
			& NDCG@50                  & 0.0418          & 0.0563            & 0.0632 & 0.0493 & 0.0756  & {\uline{0.0794}} & 0.0688 & 0.0716       & \textbf{0.0858} & 8.06\%                   \\ \hline
		\end{tabular}
	\end{table*}
	\subsection{Experimental Settings}
	\begin{table}[]
		\centering
		\caption{Summary of the datasets. The V and T represent the dimensions of visual and textual modalities, respectively.}
		\label{tab:table1}
		\scriptsize
		\setlength{\tabcolsep}{1mm}{
			\begin{tabular}{@{\extracolsep{\fill}}lcccccc@{}}
				\toprule
				Dataset & \#User & \#Item & \#Behavior & Spasity & V    & T   \\ \midrule
				Baby    & 19,445 & 7,050  & 160,792    & 99.88\% & 4,096 & 384 \\
				Video   & 24,303 & 10,672 & 231,780    & 99.91\% & 4,096 & 384 \\
				Beauty  & 22,363 & 12,101 & 198,502    & 99.93\% & 4,096 & 384 \\ \bottomrule
		\end{tabular}}
	\end{table}

	\subsubsection{Datasets}
	We conduct extensive experiments on the publicly available Amazon dataset, which includes both textual and image data. From this dataset, we select three subsets: (a) \uline{Baby}, (b) \uline{Video} Games, and (c) \uline{Beauty}. Table \ref{tab:table1} presents relevant dataset information, with data sparsity measured by the ratio of interactions to the product of the number of users and items. The preprocessed datasets incorporate both visual and textual features. Consistent with the approach in \cite{bm3,freedom}, we utilize the 4096-dimensional visual features extracted and released in \cite{Justifying}. For textual features, we generate 384-dimensional text embeddings by combining the titles, descriptions, categories, and brands of each item using sentence-transformers \cite{sentence}.
	
	\subsubsection{Implementation Details} \label{Imp}
	We conduct our experiments on an Intel Xeon Platinum 8358 machine at 2.60 GHz with Ubuntu 18.04 using an NVIDIA A100 Tensor Core GPU with 80 GB GPU memory. 
	Similar to existing works \cite{bm3, mgcn, lattice}, we standardize the embedding size of users and items to 64. The embedding parameters are initialized using the Xavier method \cite{Xavier}, and the Adam optimizer \cite{adam} is employed with a learning rate of 0.001 and a batch size of 2048. To ensure a fair comparison, we meticulously tune the parameters of each model according to their respective published papers. The proposed MGNM model is implemented using PyTorch \cite{pytorch}. All datasets are fine-tuned to align with their optimal settings.
	Specifically, we set the GNN layers to $2$, select $ \delta$ from $\left \{ 0.01,0.001,1e-4,1e-5,1e-6 \right \} $, and select $\beta$ from $\left \{ 0.1,0.01,0.001,1e-4,1e-5 \right \} $. To consider convergence, we adopt an early stopping strategy if the validation set does not improve after $5$ epochs to avoid overfitting. The original settings of the comparative methods are followed to achieve optimal performance.
	
	\subsection{Baselines}
	To evaluate the effectiveness of our proposed model, we conducted comparisons with several benchmark recommendation models. These baselines can be divided into two categories: traditional methods (relying solely on interaction data) and multimodal methods (utilizing both interaction and multimodal data):
	
	\subsubsection{Traditional Methods}
	\begin{itemize}
		\item BPR \cite{bpr}: This is a classic collaborative filtering model optimized with pairwise ranking loss using Bayesian methods.
		
		\item LightGCN \cite{LightGCN}: This is a collaborative filtering method based on GCN, simplified to perform linear propagation and aggregation between neighbors only, making it more suitable for recommendation.
	\end{itemize}
	
	\begin{table*}[t!]
		\caption{Performance Comparison under different modalities. Image and Text denote running MGNM on the visual and textual modality, respectively.}
		\centering
		\label{tab:my-table}
		\begin{tabular}{c|ccc|ccc|ccc}
			\hline
			\multirow{2}{*}{Modality} & \multicolumn{3}{c|}{Baby}              & \multicolumn{3}{c|}{Video}             & \multicolumn{3}{c}{Beauty}             \\ \cline{2-10} 
			& w/o text & w/o image & MGNM             & w/o text & w/o image & MGNM             & w/o text & w/o image & MGNM             \\ \hline
			Recall@5                  & 0.0338   & 0.0381    & \textbf{0.0412} & 0.0804   & 0.0847    & \textbf{0.0915} & 0.0624   & 0.0679    & \textbf{0.0698} \\
			Recall@10                 & 0.0530   & 0.0587    & \textbf{0.0666} & 0.1279   & 0.1308    & \textbf{0.1408} & 0.0938   & 0.1045    & \textbf{0.1051} \\
			Recall@20                 & 0.0856   & 0.0927    & \textbf{0.1003} & 0.1891   & 0.1958    & \textbf{0.2056} & 0.1359   & 0.1471    & \textbf{0.1505} \\
			Recall@50                 & 0.1485   & 0.1601    & \textbf{0.1669} & 0.2954   & 0.3040    & \textbf{0.3163} & 0.2034   & 0.2230    & \textbf{0.2236} \\ \hline
			NDCG@5                    & 0.0214   & 0.0247    & \textbf{0.0275} & 0.0543   & 0.0577    & \textbf{0.0625} & 0.0420   & 0.0461    & \textbf{0.0469} \\
			NDCG@10                   & 0.0277   & 0.0315    & \textbf{0.0358} & 0.0699   & 0.0729    & \textbf{0.0788} & 0.0525   & 0.0582    & \textbf{0.0587} \\
			NDCG@20                   & 0.0361   & 0.0402    & \textbf{0.0445} & 0.0858   & 0.0898    & \textbf{0.0957} & 0.0636   & 0.0695    & \textbf{0.0706} \\
			NDCG@50                   & 0.0488   & 0.0539    & \textbf{0.0579} & 0.1078   & 0.1122    & \textbf{0.1185} & 0.0776   & 0.0853    & \textbf{0.0858} \\ \hline
		\end{tabular}
	\end{table*}

	\subsubsection{Multimodal Methods}
	\begin{itemize}
		\item VBPR \cite{VBPR}: This classic method incorporates modality features into collaborative filtering by combining preprocessed visual features into item embeddings as representations for recommendation. It can be seen as an extension of the BPR model.
		
		\item 	MMGCN \cite{mmgcn}: This method encodes each modality using GCN to learn user preferences and then learns user and item representations from modalities through combination.
		
		\item 	DualGNN \cite{dualgnn}: This method constructs a user-user isomorphic graph from the representations learned from the user-item graph and integrates neighbor representations along the path.
		
		\item 	LATTICE \cite{lattice}: This method introduces a modality-related item-item graph and learns user and item representations by performing convolution operations on the item-item graph and user-item graph, finally aggregating information obtained from all modalities.
		
		\item 	BM3 \cite{bm3}: This method eliminates the requirement of randomly sampled negative examples through a simplified self-supervised framework and employs a dropout mechanism to perturb original embeddings of users and items.
		
		\item MGCN \cite{mgcn}: This method develops a behavior-guided purifier to avoid modality noise contamination and fuses the representations of multimodel information by the behavior-aware fuser.
		
	\end{itemize}
	\subsection{Setup and Evaluation Metrics}
	To ensure a fair comparison, we adopt the evaluation settings used in \cite{dualgnn, bm3, lgmrec}, splitting each user's interaction history into training, validation, and test sets at an 8:1:1 ratio. 
	In terms of model evaluation, we adopt two evaluation metrics i.e., Recall and NDCG, to fully verify our proposed MGNM on three datasets:
	\begin{equation}
		Recall=\frac{TP}{TP+FN},
	\end{equation}
	
	\begin{equation}
		CG_{K} =\sum_{i=1}^{K} rel_{i} ,
	\end{equation}
	
	\begin{equation}
		DCG_{K} =\sum_{i=1}^{K} \frac{2^{rel\left ( i \right ) } +1}{log_{2} \left ( i+1 \right ) } ,
	\end{equation}
	
	\begin{equation}
		NDCG=\frac{DCG}{iDCG} .
	\end{equation}
	
	We evaluate the recommendation performance of various models using Recall@K and NDCG@K metrics. In the recommendation phase, all unobserved user-item interactions are considered as unseen items. For our experiments, we empirically configure $K$ to be 5, 10, 20, and 50.
	
	\begin{table*}[ht!]
		\centering
		\caption{Hyperparameter experiments conducted on the Baby dataset.}
		\label{baby}
		\begin{tabular}{ccccccccc}
			\hline
			\textbf{$\delta $=1e-4} & Recall@5		& Recall@10       & Recall@20       & Recall@50       & NDCG@5          & NDCG@10         & NDCG@20         & NDCG@50         \\ \hline
			$\beta$=0.1             & 0.0366                          & 0.0572          & 0.0852          & 0.1422          & 0.0250          & 0.0317          & 0.0389          & 0.0505          \\
			$\beta$=0.01            & 0.0418                          & 0.0668          & 0.0984          & 0.1639          & 0.0278          & 0.0360          & 0.0441          & 0.0573          \\
			$\beta$=0.001           & 0.0410                          & 0.0662          & 0.0983          & 0.1647          & 0.0275          & 0.0358          & 0.0441          & 0.0575          \\
			\textbf{$\beta$=1e-4}   & \textbf{0.0412}                 & \textbf{0.0666} & \textbf{0.1003} & \textbf{0.1669} & \textbf{0.0275} & \textbf{0.0358} & \textbf{0.0445} & \textbf{0.0579} \\
			$\beta$=1e-5            & 0.0404                          & 0.0649          & 0.0993          & 0.1686          & 0.0269          & 0.0350          & 0.0438          & 0.0578          \\ \hline
			\textbf{$\beta$=1e-4}   & Recall@5                        & Recall@10       & Recall@20       & Recall@50       & NDCG@5          & NDCG@10         & NDCG@20         & NDCG@50         \\ \hline
			$\delta $=0.01          & 0.0399                          & 0.0642          & 0.0979          & 0.1634          & 0.0268          & 0.0347          & 0.0434          & 0.0566          \\
			$\delta $=0.001         & 0.0411                          & 0.0641          & 0.0990          & 0.1665          & 0.0274          & 0.0350          & 0.0440          & 0.0576          \\
			\textbf{$\delta $=1e-4} & \textbf{0.0412}                 & \textbf{0.0666} & \textbf{0.1003} & \textbf{0.1669} & \textbf{0.0275} & \textbf{0.0358} & \textbf{0.0445} & \textbf{0.0579} \\
			$\delta $=1e-5         & 0.0405                          & 0.0653          & 0.0991          & 0.1682          & 0.0270          & 0.0352          & 0.0439          & 0.0579          \\
			$\delta $=1e-6         & 0.0409                          & 0.0665          & 0.0995          & 0.1660          & 0.0275          & 0.0359          & 0.0444          & 0.0578          \\ \hline
		\end{tabular}
	\end{table*}
	
	\begin{table*}[ht!]
		\centering
		\caption{Hyperparameter experiments conducted on the Video dataset.}
		\label{video}
		\begin{tabular}{ccccccccc}
			\hline
			\textbf{$\delta $=1e-4} & Recall@5		& Recall@10       & Recall@20       & Recall@50       & NDCG@5          & NDCG@10         & NDCG@20         & NDCG@50         \\ \hline
			$\beta$=0.1             & 0.0878                          & 0.1353          & 0.1978          & 0.3069          & 0.0592          & 0.0748          & 0.0911          & 0.1136          \\
			$\beta$=0.01            & 0.0913                          & 0.1397          & 0.2031          & 0.3097          & 0.0624          & 0.0783          & 0.0948          & 0.1169          \\
			$\beta$=0.001           & 0.0905                          & 0.1376          & 0.2007          & 0.3126          & 0.0614          & 0.0769          & 0.0934          & 0.1164          \\
			\textbf{$\beta$=1e-4}   & \textbf{0.0915}                 & \textbf{0.1408} & \textbf{0.2056} & \textbf{0.3163} & \textbf{0.0625} & \textbf{0.0788} & \textbf{0.0957} & \textbf{0.1185} \\
			$\beta$=1e-5            & 0.0901                          & 0.1383          & 0.2030          & 0.3161          & 0.0610          & 0.0769          & 0.0938          & 0.1170          \\ \hline
			\textbf{$\beta$=1e-4}   & Recall@5                        & Recall@10       & Recall@20       & Recall@50       & NDCG@5          & NDCG@10         & NDCG@20         & NDCG@50         \\ \hline
			$\delta $=0.01          & 0.0906                          & 0.1386          & 0.2009          & 0.3097          & 0.0617          & 0.0775          & 0.0937          & 0.1162          \\
			$\delta $=0.001         & 0.0928                          & 0.1418          & 0.2041          & 0.3114          & 0.0626          & 0.0787          & 0.0950          & 0.1172          \\
			\textbf{$\delta $=1e-4} & \textbf{0.0915}                 & \textbf{0.1408} & \textbf{0.2056} & \textbf{0.3163} & \textbf{0.0625} & \textbf{0.0788} & \textbf{0.0957} & \textbf{0.1185} \\
			$\delta $=1e-5         & 0.0909                          & 0.1388          & 0.2034          & 0.3140          & 0.0618          & 0.0776          & 0.0945          & 0.1173          \\
			$\delta $=1e-6         & 0.0905                          & 0.1387          & 0.2019          & 0.3117          & 0.0610          & 0.0769          & 0.0934          & 0.1160          \\ \hline
		\end{tabular}
	\end{table*}
	
	\begin{table*}[ht!]
		\centering
		\caption{Hyperparameter experiments conducted on the Beauty dataset.}
		\label{beauty}
		\begin{tabular}{ccccccccc}
			\hline
			\textbf{$\delta $=0.01}   & Recall@5		& Recall@10       & Recall@20       & Recall@50       & NDCG@5          & NDCG@10         & NDCG@20         & NDCG@50         \\ \hline
			$\beta$=0.1             & 0.0616                          & 0.0931          & 0.1303          & 0.1973          & 0.0423          & 0.0528          & 0.0626          & 0.0766          \\
			$\beta$=0.01            & 0.0687                          & 0.1051          & 0.1492          & 0.2216          & 0.0459          & 0.0580          & 0.0697          & 0.0847          \\
			$\beta$=0.001           & 0.0680                          & 0.1014          & 0.1449          & 0.2178          & 0.0456          & 0.0567          & 0.0682          & 0.0833          \\
			\textbf{$\beta$=1e-4} & \textbf{0.0698}                 & \textbf{0.1051} & \textbf{0.1505} & \textbf{0.2236} & \textbf{0.0469} & \textbf{0.0587} & \textbf{0.0706} & \textbf{0.0858} \\
			$\beta$=1e-5         & 0.0693                          & 0.1047          & 0.1495          & 0.2240          & 0.0467          & 0.0584          & 0.0703          & 0.0857          \\ \hline
			\textbf{$\beta$=1e-4} & Recall@5                        & Recall@10       & Recall@20       & Recall@50       & NDCG@5          & NDCG@10         & NDCG@20         & NDCG@50         \\ \hline
			\textbf{$\delta $=0.01}   & \textbf{0.0698}                 & \textbf{0.1051} & \textbf{0.1505} & \textbf{0.2236} & \textbf{0.0469} & \textbf{0.0587} & \textbf{0.0706} & \textbf{0.0858} \\
			$\delta $=0.001           & 0.0684                          & 0.1032          & 0.1492          & 0.2264          & 0.0459          & 0.0575          & 0.0697          & 0.0857          \\
			$\delta $=1e-4          & 0.0686                          & 0.1054          & 0.1486          & 0.2231          & 0.0463          & 0.0585          & 0.0700          & 0.0854          \\
			$\delta $=1e-5           & 0.0682                          & 0.1054          & 0.1485          & 0.2250          & 0.0459          & 0.0583          & 0.0697          & 0.0856          \\
			$\delta $=1e-6           & 0.0675                          & 0.1040          & 0.1470          & 0.2236          & 0.0452          & 0.0574          & 0.0687          & 0.0846          \\ \hline
		\end{tabular}
	\end{table*}
	
	\subsection{Overall Performance}
	The overall comparison results on three different downstream datasets are shown in Table \ref{table2}, from which we obtain several observations.
	
	\subsubsection{Advantages of MGNM} MGNM outperforms other baseline models and exhibits good performance across different datasets. This improvement can be attributed to the use of GF, which filter out noise from the modality features, making the filtered modality information more aligned with user preferences. By eliminating the correlated user-item representations and modality-related representations, MGNM ensures that redundant information resulting from feature reuse does not affect recommendation performance.
	
	\subsubsection{Effectiveness of Multimodal Features} The introduction of rich modality information benefits user and item representations. It is evident from the experiments that even simple linear fusion of visual features like VBPR outperforms its baseline BPR in terms of performance. Moreover, multimodal models based on LightGCN (i.e., BM3, Lattice, MGCN) achieve better results than LightGCN.
	% \usepackage{multirow}
	
	\subsection{Ablation Study}
	First, we conduct an ablation analysis of the impact of each modality on recommendation performance, as shown in Table \ref{tab:my-table}. We obtain the following conclusions:
	
	\subsubsection{Multimodal Performance} Multimodal recommendation performance is significantly better than that of single modality. It's evident from the graph that the performance of MGNM is better than that of pure textual or visual modality, as user behavior is often influenced by multimodal information of items.
	
	\subsubsection{Impact of Text Modality} For the MGNM model, the text modality has a greater impact on performance. Removing text information leads to lower recommendation performance compared to removing visual information, indicating that text information plays a larger role in the recommendation process.
	
	\subsection{Hyperparameter Analysis}
	\label{hyperparameter}
	To eliminate the feature redundancy in GNNs, we utilize the DDR loss. Specifically, we conduct sensitivity analysis experiments to determine the optimal hyperparameter settings for different datasets. According to Table \ref{baby} and Table \ref{video}, the optimal settings are achieved when $\delta $ and $ \beta $ are both configured to 0.0001. Conversely, for the Beauty dataset in Table \ref{beauty}, the optimal settings are achieved when $\delta $ is configured to 0.01 and $\beta $ is configured to 0.0001. This indicates that the impact of GNN feature redundancy varies across different datasets. Additionally, it suggests that the effects of collaborative-based redundancy and modality-based redundancy are also distinct. However, a fundamental conclusion can be drawn: smaller $\delta $ and $ \beta $ promote task performance. When these values are too large, excessive removal of correlation may adversely affect recommendation performance. Conversely, when these values are too small, over-correlated features may still exist in the GNN, affecting recommendation effectiveness. Therefore, finding an appropriate level of redundancy removal is crucial for effective recommendation.
	
	\section{Conclusion}
	\label{Conclusion and Feture Work}
	In this work, we propose a model called MGNM that incorporates both local interaction and global interaction. 
	In the local interaction process, a DDR loss function is introduced to remove the feature redundancy in collaboration-related and modality-related user-item features. It calculates the correlation among the features and controls the degree of cutting the correlation with a hyperparameter. 
	Meanwhile, in the global interaction, a modality-based feature filter is designed to improve the performance of recommendations by guiding the behavioral information through the dual modality information. It filters out the noise that exists in the original behavioral information. Extensive experiments on three datasets show that our method achieves state-of-the-art results.

	\normalem
	\bibliographystyle{IEEEtran}
	\bibliography{IEEEabrv,second}
	\newpage

\end{document}